\documentclass[12pt]{article}
\usepackage[latin1]{inputenc}
\usepackage{slashed}
\usepackage{amssymb}
\usepackage{amsmath}
\usepackage{amsfonts}
\usepackage{indentfirst}
\usepackage{color}
\usepackage[top=2cm,bottom=2cm,left=3cm,right=2cm]{geometry}
\newcommand{\nin}{\noindent}
\newcommand{\be}{\begin{equation}}
\newcommand{\ee}{\end{equation}}
\newcommand{\bea}{\begin{eqnarray}}
\newcommand{\eea}{\end{eqnarray}}

\newcommand{\nn}{\nonumber\\}

\newcommand{\ol}{\overline}

\begin{document}
 
\begin{flushleft} 
KCL-PH-TH/2014-14 \quad LCTS/2014-16
\end{flushleft}
 
\vspace{1cm}
 
\begin{center} 
 
{\Large{\bf Quasi-relativistic fermions and \\ dynamical flavour oscillations} }

\vspace{1cm}

{\bf Jean Alexandre}$^{(a)}$, {\bf Julio Leite}$^{(a)}$ and {\bf Nick E. Mavromatos}$^{(a,b)}$

\vspace{0.5cm}

\small{
$(a)$ King's College London, Department of Physics, Theoretical Particle Physics and Cosmology Group, London WC2R 2LS, UK \\
$(b)$ CERN, Physics Department, Theory Division, Geneva 23 CH-1211 Switzerland}
 
\vspace{1cm}

{\bf Abstract}
 
\end{center} 
 
\nin We introduce new Lorentz-symmetry violating kinematics for a four-fermion interaction model,
where dynamical mass generation is allowed, irrespectively of the strength of the coupling. In addition, these
kinematics lead to a quasi-relativistic dispersion relation, in the sense that it 
is relativistic in both the infrared and the ultraviolet,
but not in an intermediate regime, characterized by the mass $M$. 
For two fermions, we show that a flavour-mixing mass matrix is generated dynamically, and
the Lorentz symmetric limit $M\to\infty$ leads to two free relativistic fermions, with flavour oscillations. 
This model, valid for either Dirac or Majorana fermions, can describe any set of phenomenological values for the eigen masses and the mixing angle.

\vspace{1cm}

\section{Introduction}

The generation of quark, lepton and vector boson masses, as described in the standard model due to their coupling with the 
Higgs boson, seems confirmed by the experimental results at the Large Hadron Collider \cite{Higgs}, with the discovery
of a Higgs-like scalar particle. 
However, the origin of neutrino masses is still not well established, although the seesaw mechanism seems the most elegant and 
simple for such a purpose~\cite{seesaw}, even if sterile neutrinos have not yet been discovered in Nature~\cite{white2}.

The possibility of generating neutrino masses dynamically without the involvement of heavy right-handed states is therefore still an open question,
although the dimension-5 Weinberg operator~\cite{Weinberg} provides a simple mechanism to generate neutrino masses, without introducing right-handed fields.
We consider here 
a dynamical mechanism based on Lorentz invariance violating kinematics (LIV), which contains higher-order space derivatives, 
but keeps one 
time derivative, in order not to generate ghosts. The kinematics are different than in Lifshitz-type models (see 
\cite{Lifreview} for a review), and our present mechanism involves ``quasi-relativistic'' fermions, in the sense that, in 
both the infrared (IR) 
and the ultraviolet (UV), their dispersion relation is almost relativistic, and it differs substantially from relativistic 
kinematics in an 
intermediate energy regime only, characterised by a mass scale $M$.

In addition to the implication of quasi-relativistic kinematics, our present Lagrangian allows the dynamical 
generation of mass from a four-fermion interaction, however small the coupling strength is. This is not the case for a Lorentz 
symmetric four-fermion
interaction, where a critical coupling is naturally defined by the gap equation (see for example the Nambu-Jona-Lasinio model 
\cite{NJL} - NJL). 
Lifshitz four-fermion interaction models also allow dynamical mass generation for any coupling strength
\cite{Lif4fermion}, but with a dispersion relation which implies a large deviation from relativistic kinematics in the UV. 

A non-trivial consequence of our model is the analytic properties  of the dynamical mass, as a function of the coupling constant. 
This feature is unusual, since a fermion mass cannot be generated by quantum corrections only, from a perturbative expansion in 
the Standard Model. Although we make use of a non-perturbative approach to calculate the dynamical mass, an expansion of the 
result in the coupling constant could be obtained by a one-loop calculation.

We note here other related works. In~\cite{MyersPosp}, for example, the authors identify
the LIV dimension 5 operators which lead to cubic modifications of dispersion relations for scalars, fermions and vector particles. 
More general Lorentz- and CPT-violating terms with arbitrary mass dimension have been studied and classified in~\cite{KostMewes1} 
and~\cite{KostMewes2},
in the photon and neutrino sectors, respectively. There are also studies considering perturbative and non-perturbative properties
of specific LIV higher order operators, for examples see~\cite{Mariz} and~\cite{Reyes}. Moreover, some studies regarding Lorentz violation
in the neutrino sector whether considering higher order operators or not have been performed, many of them regarding neutrino oscillations, such as
~\cite{Klink} and \cite{DiazKost}.

In section 2 we show the main properties of our model for the massive and massless single flavour case.
The possibility to generate masses dynamically for any coupling strength $g$ will be important to recover the Lorentz-symmetric 
limit when $M\to\infty$.
Indeed, the dynamical mass we find is proportional to $g^2M$, such that it can be kept fixed if we take the limits $M\to\infty$ 
and $g\to0$ simultaneously,
in such a way that $g^2M\to$ constant. Also, in this limit, the four-fermion interaction vanishes, leaving us with a free relativistic
fermion whose mass has been generated by quantum correction.
We believe that the possibility of generating a mass dynamically for all values of the coupling constant, feature which is common to Lifshitz models, 
is related to superluminality. In our case, as we explain bellow, a dispersion relation which is not superluminal would
lead to a finite gap equation which has a non-trivial solution only if the coupling constant is larger than a critical value.

Section 3 then generalizes the dynamical mass generation for two fermion flavours, where the relativistic limit $M\to\infty$
consists in two massive free fermions, with a flavour-mixing mass matrix generated dynamically. 
A similar approach is considered in \cite{ALM}, where fermions interact with a LIV Abelian gauge field, 
which plays the role of regulator and eventually decouples from fermions in the Lorentz symmetric limit. 

In section 4 we extend our results to Majorana fermions,  in both situations where either only left-handed neutrinos 
are present or additional sterile right-handed neutrinos are included. 

Conclusions and outlook are presented insection 5, and two Appendices give
technical details on {\it(i)} a microscopic gravitational model, based on the analysis of \cite{mavroLV}, which motivates our study; 
{\it(ii)} the effective action for the auxiliary field which is introduced in section 2.2.

\section{Single-flavour case}

\subsection{Massive model and classical properties}

The fermion sector of the Standard Model Extension \cite{KostMewes2} can be expressed in the form  
\be
{\cal L}_{SME}=\ol\psi\left(i\slashed\partial-m+Q\right)\psi~,
\ee
where $Q$ contains the LIV terms, and can be expanded on a basis of gamma matrices 
\be
Q=A+iB\gamma^5+C_\mu\gamma^\mu+D_\mu\gamma^\mu\gamma^5+E_{\mu\nu}\sigma^{\mu\nu}~.
\ee
In the previous expansion, $A,B,C_\mu,D_\mu,E_{\mu\nu}$ can contain any number of derivatives, including terms which are either odd or even under
the discrete symmetry CPT. The different coefficients associated to these terms can arise from vacuum expectation values (vev) of tensors of different ranks,
and should satisfy upper bounds for Lorentz symmetry violation, imposed by experiments \cite{upper}. We consider here the specific case
\be
A=\frac{b}{M}\Delta~,~~C_0=-i\frac{a}{M^2}\Delta\partial_0~,~~\vec C=-i\frac{c}{M^2}\Delta\vec\partial~,~~B=D_\mu=E_{\mu\nu}=0~,
\ee
where $a,b,c$ are dimensionless constants ($a>0$ and $c>0$), such that
\be
Q=-i\partial_0\gamma^0\frac{a}{M^2}\Delta+i\vec\partial\cdot\vec\gamma \left(i\frac{b}{M}\vec\partial\cdot\vec\gamma+\frac{c}{M^2}\Delta\right)~.
\ee
This choice is motivated by a gravitational microscopic model, based on the low-energy limit of a string theory on a three brane universe, which is embedded, 
from an effective three-brane observer view point, in a bulk space-time punctured with point-like defects (\emph{D-particles}) \cite{mavroLV}. 
As shown in Appendix A, this model provides an elegant construction of the LIV operator $Q$ from a fundamental theory, 
which allows the generation of the operator $Q$ in a natural way. 
The Lorentz symmetric limit $M \to \infty$, which we shall consider in this work, follows, in this microscopic context, when the density of D-particles becomes vanishingly small.  
We note that similar steps also lead to the minimal LIV Abelian 
gauge model of \cite{alexandre}. 

Our model is then defined by the Lagrangian 
\be\label{model}
\mathcal{L}_1 = \bar{\psi} \left[i\partial_0\gamma^0\left(1-\frac{a}{M^2}\Delta\right)
-i\vec\partial\cdot\vec\gamma\left(1-i\frac{b}{M}\vec\partial\cdot\vec\gamma-\frac{c}{M^2}\Delta\right) -m\right]\psi 
 + \frac{g^2}{M^2} (\ol\psi\psi)^2,
\ee
where $g^2$ is a dimensionless coupling, such that the mass scale $M$ is used both to control the LIV scale and the 
strength of the four-fermion interaction. 
As we explain bellow, the relevant case for our study corresponds to non-vanishing coefficients $a,b,c$, 
which will lead to a quasi-relativistic dispersion relation and the absence of critical coupling for the generation of dynamical mass. 

In what follows, we will assume that $m<<M$. The dispersion relation for the Lagrangian (\ref{model}) is
\be\label{disprel}
\omega^2=m^2\left(\frac{1+bp^2/(Mm)}{1+ap^2/M^2}\right)^2+p^2\left(\frac{1+cp^2/M^2}{1+ap^2/M^2}\right)^2~.
\ee
For all the values of $a,b,c$, the Lorentz symmetric limit is naturally obtained for $M\to\infty$, at fixed $p$ and $m$.
Few specific cases are worth mentioning:\\

$\bullet$ {\it $a=0$ and $c\ne0$:} In this case the Lagrangian (\ref{model}) is equivalent to a $z=3$ Lifshitz theory, where time has been rescaled by $M^2$,
and $\omega^2\simeq p^6/M^4$ in the UV. We do not consider this situation here, since we wish to avoid such a deviation from relativistic kinematics in the UV;\\

$\bullet$ {\it $a\ne0$ and $b=c=0$:} The energy is a decreasing function of the momentum for $p^2>M^2$, which is an unphysical situation we will discard;\\

{$\bullet$ \it $a\ne0$, $b\ne0$ and $c=0$:} Here the energy goes to a constant value in the UV $\omega^2\simeq (bM/a)^2$. We also discard this possibility,
since it leads to group velocity which goes to 0 when $p\to\infty$;\\

$\bullet$ {\it $a\ne0$ and $c\ne0$:} In this situation, the UV regime is relativistic since $\omega^2\simeq (cp/a)^2$ and only the intermediate regime 
$p\sim M$ is not relativistic. This is the situation this article focuses on.\\

\nin If we impose $\omega$ to be an increasing function of $p$, the different constants in the model (\ref{model}) must satisfy
\be
2b^2+4c\ge a+2ab\, m/M~,
\ee
and without loss of generality\footnote{A more general study would keep free parameters $a,b,c$, but our aim here is to
give emphasis on the mechanism of dynamical mass generation, for which the choice $a=c=1$ is enough.} 
we shall choose $a=c=1$. The product of the group and phase velocities is then 
\be\label{vpvg}
v_pv_g=\frac{\omega}{p}\frac{d\omega}{dp}=1+\frac{2}{M^2}\frac{m+bp^2/M}{(1+p^2/M^2)^3}(bM-m)~.
\ee
The latter expression shows that, for a typical Standard Model mass $m$ and a typical Planckian mass $M$, 
the upper bound for Lorentz symmetry violation \cite{upper}
\be\label{upperbound}
|v_pv_g-1|\lesssim10^{-16} ~,
\ee
is satisfied for $p\lesssim 10^{-8}M$ (if $b$ is of order 1), which although far outside the current range of energies 
available experimentally in the laboratory, approaches the Greisen-Zatsepin-Kuzmin cut off limit of high energy cosmic rays. 
However, as shown in the microscopic model of Appendix A,  the scale $M$ is inversely proportional 
to the density of defects in the quantum-gravity ``foam'' medium, and as such, for low density media the scale $M$ may be 
considerably greater than the Planck scale, thereby avoiding any phenomenological constraints. \\
Finally, the dispersion relation (\ref{disprel}) is similar to 
a resummation of higher-order powers of the momentum $p$, although the Lagrangian (\ref{model}) is local and contains a finite 
number of space derivatives.

The bare propagator $S$ for the model (\ref{model}) is 
\be\label{prop}
S=i~\frac{(\omega \gamma^0 -\vec{p}\cdot\vec{\gamma})(1+p^2/M^2)+m+bp^2/M}
{(\omega^2-p^2)(1+p^2/M^2)^2-(m+bp^2/M)^2}~,
\ee
and we note that, if $b\ne0$, its trace does not vanish in the massless case $m=0$, which will be important 
for the analytical properties of the dynamical mass, as explained bellow.

\subsection{Massless model and dynamical mass generation}

We study here the possibility to generate a mass dynamically, in the situation where the bare mass $m$ vanishes. 
We consider the usual approach which consists in introducing a Yukawa coupling of fermions to an auxiliary field $\phi$, integrate over
fermions, and look for a non-trivial minimum for the effective potential $V(\phi)$, which leads to a mass term in the original
Yukawa interaction. This approach neglects fluctuations of the auxiliary field about its vev, 
but these can be omitted when $g^2\to0$ that we consider for the Lorentz symmetric limit (see next subsection).

We consider then the intermediate Lagrangian
\be\label{intermediate}
\mathcal{L}_1'= \bar{\psi} \left[i(\partial_0\gamma^0-\vec\partial\cdot\vec\gamma)\left(1-\frac{\Delta}{M^2}\right)
+b\frac{\Delta}{M}\right]\psi -\frac{M^2}{4}\phi^2-g\phi\ol\psi\psi~,
\ee
for which the integration over $\phi$ leads back to the original model (\ref{model}) with $a=c=1$. 
Note that the auxiliary field does not propagate at the tree level, and its large mass is an important feature of this approach to dynamical mass generation,
because it suppresses fluctuations of $\phi$ about its vev $\phi_1$, such that $g\phi\simeq g\phi_1$ can be identified with the fermion mass.
As a consequence, it is sufficient to consider a homogeneous configuration for $\phi$, in order to 
calculate the effective potential $V(\phi)$ and its minimum $\phi_1$.
Nevertheless, the field $\phi$ can be physically interpreted as a scalar collective excitation
of the original fermionic fundamental degrees of freedom, whose kinetic term is generated by integrating out fermions, if one allows $\phi$ to 
depend on spacetime coordinates. We derive this
kinetic term in Appendix B, where we show there that it vanishes in the Lorentz-symmetric limit considered in the next subsection, which 
is consistent with the assumption that $\phi$ is frozen to its vev $\phi_1$ in this limit.

From the Lagrangian (\ref{intermediate}), 
the integration over fermions for a homogeneous field $\phi$ leads to the effective potential
\be\label{Veff1}
V_1(\phi)= \frac{M^2}{4}\phi^2 + i~ tr \int \frac{d^4 p}{(2 \pi)^4} 
\ln\left[(\omega\gamma^0-\vec p\cdot\vec\gamma)(1+p^2/M^2)-bp^2/M-g\phi\right]~.
\ee
The gap equation is obtained from the minimization of the potential $(dV_1/d\phi)_{\phi_1}=0$, which, after a Wick rotation, gives
\be\label{minimize}
\frac{M^2}{2}\phi_1=
\frac{g}{\pi^3}\int p^2 dp \int d\omega \left[\frac{ (g\phi_1 + bp^2/M)}{(\omega^2+p^2)(1+p^2/M^2)^2+(g\phi_1+bp^2/M)^2} \right]~,
\ee
and leads to the dynamical mass $m_{dyn}=g\phi_1$.
The integration over frequencies in eq.(\ref{minimize}) leads to 
\be\label{gap}
\mu\frac{\pi^2}{2g^2}=\int \frac{x^2dx~(\mu+bx^2)}{(1+x^2)\sqrt{x^2(1+x^2)^2+(\mu+bx^2)^2}}~,
\ee
where $x=p/M$ and $\mu = m_{dyn}/M$. We note that, if $b\ne0$, the gap equation (\ref{gap}) does not have a vanishing solution 
$\mu=0$ unlike the case in conventional studies of dynamical mass generation. Also,
if we consider $b=0$, then the remaining integral in the gap equation (\ref{gap}) is convergent, and the existence of a non-vanishing dynamical mass requires
the coupling constant $g$ to be larger than some critical coupling (we note that $b=0$ also coincides with a subluminal product $v_pv_g$ in eq.(\ref{vpvg})).
We disregard this possibility, since we eventually will take $g^2\to0$
for the Lorentz-symmetric limit. We therefore consider from now on $b=1$, and regularize the gap equation (\ref{gap})
by $M$, such that the domain of integration in the gap equation is $0\leq x\leq 1$.\\
An expansion of eq.(\ref{gap}) in $\mu$ gives then
\be\label{mdyn}
\mu=\frac{m_{dyn}}{M}=\frac{\alpha g^2}{1-2g^2/(5\pi^2)}+{\cal O}(\mu^2)~,
\ee
where
\be\label{alpha}
\alpha=\frac{\ln(1+2/\sqrt5)-\arctan(1/2)}{\pi^2}\simeq~0.018~.
\ee
Taking into account $g^2<<1$, we obtain finally
\be\label{mdyn2}
m_{dyn}\simeq \alpha g^2M~.
\ee
We checked that the solution (\ref{mdyn}) indeed corresponds to a minimum of the effective potential (\ref{Veff1}). 
An interesting point to note is that
the dynamical mass (\ref{mdyn}) is analytic in the coupling constant $g^2$, unlike the situation of Lifshitz 4-fermion interaction \cite{Lif4fermion,ABH},
where a dynamical mass has the typical non-analytic form 
\be\label{non-analytic}
m_{dyn}^{Lif}\simeq M\exp(-a/g^2)~, 
\ee
where $a$ is a constant. We note however that the expression (\ref{mdyn}) consists of a resummation in powers of $g^2$ and 
goes beyond a one-loop 
calculation. Nevertheless, the approximate result (\ref{mdyn2}) can be obtained from the usual one-loop correction to the fermion mass. 
This feature
is specific to the LIV propagator (\ref{prop}), whose trace does not vanish, even in the massless case. 
We are therefore
in the unusual situation where a fermion mass generated dynamically can be derived using a perturbative expansion, 
whereas a mass of the form (\ref{non-analytic}) can be obtained from a non-perturbative approach only.

For completeness, we give the expression for the dynamical mass when $b\ne1$:\\
\bea
m_{dyn}&\simeq& b g^2 M \frac{2 \ln 2 -1}{2 \pi^2} ~~~~\mbox{for}~~0\ne b<<1\\
m_{dyn}&\simeq& g^2 M \left(\frac{4-\pi}{2\pi^2} +\mathcal{O}(1/b^2)\right)~~~~\mbox{for}~~b>>1~.\nonumber
\eea 
We note that the limit $b\to0$ continuously leads to the vanishing solution 
of the gap equation (\ref{gap}) when $b=0$, as in a first order phase transition,
and the non-trivial solution (involving a critical coupling) is not recovered. In addition, the situation $b<<1$ 
leads to a suppression of LIV effects in the dispersion relation (\ref{disprel}), and thus might be more relevant than the case $b>>1$. 
The limit $b<<1$ is also supported by the microscopic model described in Appendix A, in which $b = \frac{m}{M} \tilde b $, 
with $\tilde b$ an arbitrary real number.
In the limit $m \ll M$ we are considering throughout, natural values of $\tilde b$ imply the case $b \ll 1$.  
Moreover, in this limit, the physical irrelevance of the precise numerical value of the constant $b$, as long as $b \ne 0$, 
becomes evident by the fact that it can be absorbed in a redefinition of the coupling $g$ which is taken eventually to zero. 
This is in agreement with $b$ being attributed to quantum ordering ambiguities in the microscopic model.

\subsection{Lorentz symmetric limit} 

An important point, which differs from the NJL model for example, is that 
dynamical mass generation occurs here for any coupling strength, and no critical coupling exists, 
bellow which this non-perturbative process does not occur. 
This feature allows us to take the Lorentz symmetric limit of the model, $M\to\infty$, in such a way 
that the dynamical mass (\ref{mdyn}) remains finite, provided that $g$ depends on $M$ as
\be\label{gpropto}
g(M)\sim\sqrt\frac{m_{dyn}}{\alpha M}~~,~~~~\mbox{when}~~M\to\infty~,
\ee
where $m_{dyn}$ is fixed.\\
In this simultaneous limit, where the product $Mg^2$ goes to a finite value, we are left 
with free relativistic massive fermions, for which the mass has been generated by quantum corrections. 
A similar limit is considered in \cite{ALM}, where the 
dynamical mass has the form (\ref{non-analytic}) though.

Because the model (\ref{model}) violates Lorentz symmetry, space and time derivatives are dressed differently 
by quantum corrections, which is problematic when several species interact \cite{deviation}, as we consider in the next section
\footnote{If only one kind of particle self-interacts, then one can always rescale momentum in order to absorb quantum corrections in such a way
that the dressed dispersion relation remains relativistic in the IR.}. 
Indeed, it has been shown, in different Lifshitz models for example, that loop corrections to classical 
dispersion relations lead to worse
deviations from Relativity than the classical dispersion relation \cite{alexandre,ABH,deviation,AB}. 
But since a consistent Lorentz symmetric limit in our case implies $g^2\to0$, loop corrections to the kinetic terms in the model 
(\ref{model}) eventually vanish in this limit,
such that the classical upper bound (\ref{upperbound}) for Lorentz violation remains satisfied.

To illustrate this point, let us calculate the superficial degree of divergence $D$ of an $L$-loop graph $\Sigma^{(L)}$ contributing to the self energy.
Each loop integral measure carries the mass dimension 4 and each propagator (\ref{prop}) has mass dimension -1. The corrections $p^2/M^2$ are at most 
equal to 1, since integrals are regularized by $M$: they do not play a role for the superficial degree of divergence. 
Hence $D=4L-I$, where $I$ is the number of internal propagators. 
Momentum conservation leads to $L=I-V+1$, where $V$ is the number of vertices. Also, 
since each vertex has 4 legs, and each internal propagator relates two vertices, we have $4V=E+2I$, where $E$ is the number of external propagators. 
As a consequence, we have, as for the usual NJL model,
\be
D=2L+2-\frac{E}{2}~,~~\mbox{and}~~~V=L-1+\frac{E}{2}~.
\ee
In our case though, each vertex brings a factor $g^2/M^2$, hence for the self energy ($E=2$) we have 
\be
\Sigma^{(L)}\propto\left(\frac{g^2}{M^2}\right)^VM^{D}=Mg^{2L}~.
\ee
Taking into account the limit (\ref{gpropto}), we finally obtain
\be\label{sigmaL}
\Sigma^{(L)}\propto\frac{m_{dyn}^L}{M^{L-1}}~.
\ee
The first non-trivial loop corrections to the kinetic terms occur at two loops, since the one-loop self energy is independent of the external momentum.
As a consequence we are interested in $L\ge2$, and for a {\it fixed} dynamical mass $m_{dyn}$, the property (\ref{sigmaL}) therefore shows that the 
loop correction $\Sigma^{(L)}$ goes to 0 when $M\to\infty$: quantum corrections to the kinetic terms vanish in the Lorentz symmetric limit (\ref{gpropto}).

\section{Two-flavour case and dynamical flavour oscillations}

We extend here our model to a massless fermion doublet $\Psi$, which is self interacting via the flavour-mixing coupling matrix $\tau$ 
\be\label{modelbis}
\mathcal{L}_2 = \bar{\Psi} \left[i(\partial_0\gamma^0-\vec\partial\cdot\vec\gamma)\left(1-\frac{\Delta}{M^2}\right)
+\frac{\Delta}{M}\right] \Psi + \frac{1}{M^2} (\ol\Psi\tau\Psi)^2,
\ee
where 
\be
\Psi=\begin{pmatrix} 
       \psi_1\\
       \psi_2
       \end{pmatrix}~,
~~~~\mbox{and}~~~~
\tau=\begin{pmatrix} 
       g_1 & g_3\\
       g_3 & g_2
       \end{pmatrix}~,
\ee
and we show that flavour oscillations are generated dynamically.

\subsection{Minimization of the effective potential}

The Lagrangian containing the auxiliary field, equivalent to the original Lagrangian (\ref{modelbis}), is
\be\label{lag2}
\mathcal{L}_2' = \bar{\Psi}\left[i(\partial_0\gamma^0-\vec\partial\cdot\vec\gamma)\left(1-\frac{\Delta}{M^2}\right)
+\frac{\Delta}{M}\right]\Psi-\frac{M^2}{4}\phi^2-\phi\ol\Psi\tau\Psi~,
\ee
and, in order to integrate over fermions, we should first find the eigen values, in flavour space, of the operator
\be
{\cal O}=\begin{pmatrix} 
       (\omega\gamma^0-\vec p\cdot\vec\gamma)(1+\frac{p^2}{M^2})-\frac{p^2}{M}-g_1\phi & -g_3\phi\\
       -g_3\phi & (\omega\gamma^0-\vec p\cdot\vec\gamma)(1+\frac{p^2}{M^2})-\frac{p^2}{M}-g_2\phi
       \end{pmatrix}~.
\ee
These are
\be
\lambda_\pm=(\omega\gamma^0-\vec p\cdot\vec\gamma)(1+p^2/M^2)-p^2/M-h_\pm\phi~,
\ee
where the eigen values $h_\pm$ of the coupling matrix $\tau$ are given by
\be
h_\pm=\frac{1}{2}(g_1+g_2)\pm\frac{1}{2}\sqrt{(g_1-g_2)^2+4g_3^2}~.
\ee
The effective potential for the auxiliary field is therefore
\be
V_2=\frac{M^2}{4}\phi^2+i ~tr\int \frac{d^4 p}{(2 \pi)^4} (\ln\lambda_++\ln\lambda_-)~,
\ee
and its minimization $(dV_2/d\phi)_{\phi_2}=0$ leads to 
\be
\frac{M^2}{2}\phi_2=
\sum_{s=+,-}\frac{h_s}{\pi^3}\int p^2 dp \int d\omega \left[\frac{ (h_s\phi_2 + p^2/M)}{(\omega^2+p^2)(1+p^2/M^2)^2+(h_s\phi_2+p^2/M)^2} \right]~.
\ee
The integration over frequencies leads to the following gap equation, regularized by the mass scale $M$,
\be\label{gap3}
\kappa\frac{\pi^2}{2} = \sum_{s =+,-}h_s\int_0^1 \frac{x^2dx~(h_s\kappa+x^2)}{(1+x^2)\sqrt{x^2(1+x^2)^2+(h_s\kappa+x^2)^2}} ~,
\ee
where $x = p/M $ and $\kappa = \phi_2/M$. An expansion in $\kappa$ gives
\be
\kappa = \frac{\alpha(h_+ + h_- )}{1-2(h_+^2 + h_-^2)/(5\pi^2)}+{\cal O}(\kappa^2)~,
\ee
where $\alpha$ is given by eq.(\ref{alpha}). Taking into account $h_\pm<<1$ we finally obtain 
\be\label{min2}
\kappa \simeq \alpha(g_1 + g_2 )~.
\ee
This value for the minimum of the potential gives the dynamical mass matrix, as we see in the next subsection.

\subsection{Dynamical flavour oscillations}

From the previous results, the mass matrix $\mathcal{M} = \kappa M \tau$ dynamically generated is
\be
\mathcal{M}=\alpha(g_1+g_2)M\begin{pmatrix} g_1 & g_3 \\ g_3 & g_2\end{pmatrix}~,
\ee
such that the mass eigen values $m_\pm = \kappa M h_\pm$ and the mixing angle $\theta$ are given by
\bea\label{md12}
m_\pm &=& \frac{\alpha}{2} M\left[(g_1+g_2)^2\pm\sqrt{(g_1^2-g_2^2)^2+4g_3^2(g_1+g_2)^2}\right]\nn
\tan\theta&=&\frac{g_1-g_2}{2g_3}+\sqrt{1+\left(\frac{g_1-g_2}{2g_3}\right)^2}~.
\eea
From this we can express the dimensionless couplings $g_i$ as
\bea\label{g1g2g3}
g_1 &=&\frac{\mu_++\mu_-+(\mu_+-\mu_-)\cos(2\theta)}{2\sqrt{\alpha(\mu_++\mu_-)}}~,\\
g_2 &=&\frac{\mu_++\mu_--(\mu_+-\mu_-)\cos(2\theta)}{2\sqrt{\alpha(\mu_++\mu_-)}}~,\nn
g_3 &=&\frac{\mu_--\mu_+}{2\sqrt{\alpha(\mu_++\mu_-)}}\sin(2\theta)~,\nonumber
\eea
where 
\be 
\mu_\pm = \frac{m_\pm}{M}~.
\ee
Therefore one can write the couplings $g_i$ in the form
\be\label{gi}
g_i=\frac{a_i}{\sqrt M}~~,~i=1,2,3~,
\ee
where the constants $a_i$ are completely fixed by the experimental values for $m_\pm$ and $\theta$. This expression shows the explicit 
dependence of the couplings $g_i$ on the scale $M$, for the Lorentz symmetric limit $M\to\infty$ to be taken, in such a way that we are left with
two relativistic free fermions, for which flavour oscillations have been generated dynamically. Therefore any set of values for $m_\pm$ and $\theta$
can be described by the Lorentz-symmetric limit of our model, by considering the coupling constants (\ref{g1g2g3}).

We can now discuss the oscillation probability, {\it i.e.} the probability for a neutrino to change from one flavour,
say, $\nu_{\beta_1}$ into another $\nu_{\beta_2}$, during its propagation \emph{in 
vacuo}. Generically, for the two flavour case, it is given by the following
expression~\cite{bilenky}:
\be\label{oscprob}
\mathcal{P}(\nu_{\beta_1} \to \nu_{\beta_2} ) = \sin^2(2 \theta) \sin^2\left[\frac{ (E_+ - E_-)~t}{2}\right],
\ee
where $t$ is the time in which the neutrino arrives at the detector after leaving the source at $t=0$ and $E_\pm$ are the energy of
each neutrino mass eigenstate. In the usual, Lorentz invariant case, $E_\pm \sim p + m_\pm^2/2p + \dots  $, for
$m_\pm \ll p$, with $p$ the momentum of the relativistic neutrino, so one derives the standard 
relativistic result 
\be\label{oscprob2}
\mathcal{P}(\nu_{\beta_1} \to \nu_{\beta_2} ) = \sin^2(2 \theta) \sin^2\left[\frac{(m_+^2 - m_-^2) L}{4 E} \right],
\ee
where with  $E\simeq p$ and $t\simeq L$ for relativistically fast neutrinos, $L$ being the distance between the source and the detector.

In our case, using the dispersion relation~(\ref{disprel}) and considering, as usual, $m_\pm^2/p^2\ll 1$
and $m_\pm/M<<1$, we find
\be\label{endiff}
(E_+ - E_-) t = \frac{(m_+^2 - m_-^2) L}{2 E} + (m_+ - m_-) \frac{E L}{M}+ \mathcal{O}(m_\pm^2/M^2)~.
\ee
Therefore, the corresponding oscillation probability can be written as
\bea\label{proba}
\mathcal{P}(\nu_{\beta_1} \to \nu_{\beta_2} ) &=& \sin^2(2 \theta) \sin^2\left[\frac{ (m_+^2 - m_-^2) L}{4 E} + 
(m_+ - m_-)\frac{E L}{2 M}+ ...\right]~, \nonumber \\
&\simeq& \frac{\sin^2 [A ~(g_1 + g_2)^3 \sqrt{(g_2-g_1)^2 + 4g_3^2}~]}{1 + (g_2 - g_1)^2/(4 g_3^2)},
~~~~\mbox{with}~A = \frac{\alpha^2 M^2 L}{4 E}~.
\eea
where we took into account that, on account of (\ref{md12}) and (\ref{g1g2g3}), ${\rm sin}^{-2}(2\theta)$
is just 
the denominator of the right-hand-side of  (\ref{proba}). 
According to eq.(\ref{gi}), the argument of the sine in the middle equation (\ref{proba}) goes to a finite limit when $M\to\infty$, 
since it is proportional to the finite $M^2g_i^4$, $i=1,2,3$, and so does the denominator in the expression of  the right-hand-side of (\ref{proba}). 
We also note that the first term on the argument of the sine in the middle equation (\ref{proba}) is the usual relativistic expression,
while the second term is the first
contribution coming from the Lorentz-violating features of our model. Such a second term goes to zero when $M\to \infty$, and (\ref{proba}) reduces to  
the usual relativistic oscillation probability (\ref{oscprob2}) \emph{in} Lorentz-invariant \emph{vacuo}, as expected. 
On the other hand, for finite $M$, \emph{e.g}. the case of the microscopic string model of Appendix A, the second term in~(\ref{endiff}), 
linearly suppressed by $M$, may have phenomenological 
consequences, for $M $ as large as Planck mass, $M_{\rm Pl}$,
as discussed in \cite{brustein}.

\section{Majorana fermions}

Although there is no certainty whether neutrinos are Dirac or Majorana particles, it is
likely that they are Majorana fermions. Based on this possibility, we present,
in the following sections, two ways of extending the previous results to the case of Majorana fermions, specifically neutrinos.

\subsection{Left-handed Majorana fermions}

A Majorana fermion obeys the following relation
\be
(\nu^M)^c \equiv C(\bar{\nu}^M)^T =  \nu^M~,
\ee
known as the Majorana condition, where $C$ is the charge conjugation operator \cite{bilenky}. As a consequence, the chiral components of a Majorana
field are not independent as for Dirac fields, actually, if $\nu_L$ is the left-handed component of a Majorana field, then its right-handed
component is simply given by $\nu_R = (\nu_L)^c$. Moreover, from these fields it is possible to construct a Majorana mass term, which for two flavours,
has the following form
\bea\label{mmt}
\mathcal{L}_{mass}^M = -\frac{1}{2} \bar{\nu}_L~M^M~(\nu_L)^c + h.c.~, ~~~\nu_L = \begin{pmatrix} \nu_{\beta_1 L} \\ \nu_{\beta_2 L} \end{pmatrix}~,
\eea
where $M^M$ is a $2\times 2$ symmetric mass matrix and $\beta_i$ represents the flavour of the field. On the other hand, the flavour eigenstates can be
transformed into mass eigenstates by using a unitary mixing matrix $U$, so that the Majorana mass term above is then diagonal and we find
\be\label{mterm}
\mathcal{L}_{mass}^M = -\frac{1}{2} \sum_{j=1,2} m_j~\bar{\nu}_j\nu_j~,
\ee
where $\nu_i = \nu_{i L}+(\nu_{i L})^c$ is the Majorana field. In a similar way, we can express the kinetic term of~(\ref{modelbis}) when
$\Psi \to \nu_L$ in terms of the mass eigenstates $\nu_j$, such that
\bea\label{kin}
\mathcal{L}^M_{kin} &=& \bar{\nu}_L \left[i(\partial_0 \gamma^0-\vec{\partial}\cdot \vec{\gamma}) 
\left(1- \frac{\Delta}{M^2}\right) + \frac{\Delta}{M}\right]~\nu_L\\
&=&  \frac{1}{2} \sum_{j=1,2} \bar{\nu}_j \left[i(\partial_0 \gamma^0-\vec{\partial}\cdot \vec{\gamma}) 
\left(1- \frac{\Delta}{M^2}\right) + \frac{\Delta}{M}\right]~\nu_j~,\nonumber
\eea
From now on, we consider the model given by~(\ref{modelbis}) with $\Psi \to \nu_L$ written in terms of the mass eigenstates, so that
the kinetic term is given by~(\ref{kin}). In addition, we assume that the matrix $\tau$ in~(\ref{modelbis}) is diagonal, {\it i.e.} $g_3=0$.
Thus, the calculations follow in the same way as in the previous section, with the only difference of factor $1/2$
in the kinetic term (\ref{kin}), which leads to the minimum
\be
\kappa' \simeq 2 \alpha (g_1+g_2)~.
\ee
and the following masses for left-handed Majorana fermions are generated
\be
m_i = 2 \alpha M (g_1+g_2) g_i, ~~~ i = 1,2~.
\ee
Moreover, since $g_3 =0$, no term presenting a mixture of mass eigenstates is generated. However, once we want to express this solution in terms
of the flavour eigenstates which are coupled to the $SU(2)_L$ gauge field of the standard model, flavour mixing will be
naturally generated and oscillations will be allowed.

\subsection{Seesaw-type extension}

We consider here the existence of right-handed sterile fermions (neutrinos), in addition
to left-handed active ones, and provide a seesaw-type extension of our original model.

The main idea is to consider the model given in eq.(\ref{modelbis}), however, by choosing the fermion doublet $\Psi$ to represent a left-handed ($v_L$)
and a right-handed Majorana field ($N_R$) instead of two Dirac fields or two left-handed Majorana fields, as in the previous sections. Thus, in this case,
we are working with one generation only. This configuration allows us to construct two different kinds of mass terms: 
\be\label{MD}
{\mathcal L}^{M+D}= -\frac{1}{2} {\overline \nu}_L \, m_L \, (\nu_L)^c - \overline{\nu}_L \, m^D \, N_R  -  
\frac{1}{2} {\overline N}_R \, m_R \, (N_R)^c + {\rm h.c.}~,
\ee
where $m_{L,R}$ are Majorana mass terms and $m_D$ is the usual Dirac mass term.

The solutions of this case should be the same as the ones we found for the original model with Dirac fermions,
however, if we consider $g_1 = 0$ and $g_3\ll g_2$, the mass matrix below is generated
\be
\mathcal{M} = \alpha M g_2 \begin{pmatrix} 0 & g_3 \\ g_3 & g_2 
                            \end{pmatrix}
	= \begin{pmatrix} m_L & m_D \\ m_D & m_R
\end{pmatrix}~,
\ee
with the following eigenvalues
\bea
m_{+} &\simeq& \alpha M g_2^2 = m_R\\
m_{-} &\simeq& \alpha M g_3^2 = \frac{m_D^2}{m_R} \ll m_R~.
\eea
These results imply that the heavier the right-handed fermion, the lighter the left-handed one, as in seesaw-type mechanisms.

In the original seesaw mechanism the mass term $m_D$ is generated via the Higgs mechanism, while the term $m_R$ is generated
via an unknown (non-standard model) mechanism. In the case described here, the Higgs mechanism is not needed to generate the
Dirac mass term $m_D$ (although, it can also be easily included in the model) and $m_D$ as well as $m_R$ is generated by the
mechanism described in this paper.

\section{Conclusion}

In this work we have shown that flavour oscillations among neutral fermions, representing neutrinos, can arise dynamically from 
LIV operators, and that the Lorentz symmetric limit can be recovered 
consistently, keeping dynamical masses as finite IR effects. We stress here the importance of the procedure: starting from a finite LIV mass scale $M$
and finite four-fermion coupling $g^2/M^2$, we study the dynamical generation of masses, to eventually consider the simultaneous limits 
$M \to \infty$ and $g \to 0$, in such a way that the dynamical masses remain finite. The original LIV Lagrangian can therefore be interpreted as a 
\emph{regularized} model from which masses are generated, after the regularization is removed consistently. 
The originality of our model therefore 
consists in generating flavour oscillations from quantum corrections, and not tree-level processes.
These corrections imply finite effects in the IR, even after removing the original LIV regulator, in order to 
recover the Lorentz-symmetric limit.

An additional remark related to the structure of the Standard Model is the following.
One could think that the introduction of such LIV terms in the neutrino sector can somehow bring unwanted consequences for the charged leptons.
But it is important to note that the higher-order derivative terms added in the neutrino sector are not invariant under $U(1)$ gauge transformations,
unless one introduces new interactions which are not renormalizable, such that then these LIV terms are not allowed in the charged lepton sector.
Thus, the present model does not directly imply new physics for charged leptons. Nevertheless, one can still expect radiative corrections on charged
leptons, because of their interactions with neutrinos through Weak gauge bosons. But such new effects are suppressed by the mass scale $M$, 
which eventually is taken to infinity. Therefore, it needs to be emphasized that the only observable effect of the present Lorentz violating 
model is the dynamical generation of neutrino oscillations.

An extension of the present work consists in deriving in detail the four-fermion interaction from a microscopic 
LIV  gravitational model. This interaction might be generated by the D-particle microscopic model described in Appendix A, but also from an alternative
theory to General Relativity, which breaks 4-dimensional diffeomorphism symmetry, as Horava-Lifshitz Gravity for example
\cite{HL}. The corresponding detailed phenomenology of such models remains to  be seen.

We close by stressing again that, at present, the origin of neutrino masses and their hierarchy is not known, despite the existence of compelling 
theoretical scenarios, such as the seesaw mechanism. It is therefore not impossible that dynamical phenomena, like the ones discussed in this work
and possibly originating from a more fundamental gravitational theory,
are responsible for the tiny neutrino masses and the associated hierarchies and  oscillations.

\section*{Acknowledgements}

The work  of J. L. is supported by the National Council for Scientific and
Technological Development (CNPq - Brazil), while that of 
N.E.M. is supported in part by the London Centre for Terauniverse Studies (LCTS), using funding from the European Research Council via 
the Advanced Investigator Grant 267352 and by STFC (UK) under the research grant ST/J002798/1.

\section*{Appendix A:\\ Microscopic motivation for the model}

Consider the low-energy limit of a string theory on a three brane universe, which is embedded, 
from an effective three-brane observer view point, in a bulk space-time
punctured with point-like (\emph{D-particles}). For the purposes of this work we shall concentrate on 
the interaction of relativistic (Dirac or Majorana) fermions, represented by open string excitations attached on the brane world, 
with such a background. We consider a collection of such D-particles, with $n^\star$ defects per four-dimensional string volume. 
The background acts as a microscopic regulator, eventually we shall take the density $n^\star \to 0$. 

 The interactions of particle probes with such a `medium', 
can be simply described (on average, over a collection of D-particles) by an induced target space metric, which 
deviates from the Minkowski metric by terms of the form~\cite{mavroLV}
\bea\label{recoil}
\delta g_{0i} = u_i = g_s \frac{\Delta p_i}{M_s} = g_s n^\star \frac{r\, p_i}{M_s} \ll 1, 
\eea
where $u_i$ is the recoil velocity of the D-particle, during its individual scattering with the fermion and 
$\Delta p_i = n^\star \, r\, p_i$, with $|r n^\star | < 1$ a dimensionless numerical coefficient, proportional to the D-particle density, 
which represents an average momentum transfer during the individual 
scatterings of the open string with the medium of D-particles with density $n^\star$.  The induced metric is therefore of 
Finsler type, as it depends on momenta. 
This implies modified dispersion relations for the low-energy 
excitations on the brane universe, with LIV modifications suppressed by an appropriate inverse power of the `effective mass scale'' 
\begin{equation}\label{star}
M = \frac{M_s}{g_s n^\star r}
\end{equation}
where $M_s/g_s$ is the mass of the D-particle defects, $M_s$ is the string scale (which may be viewed as the `quantum gravity scale'), 
and $g_s < 1$ is the weak string coupling.

For the case of fermions, at tree-level in string theory, that is considering open strings (representing excitations on the brane universe) 
propagating on a world-sheet with the topology of a disc, the detailed analysis of the low-energy effective field theory in the D-particle 
medium has been performed in 
\cite{volkov}, with the result that the kinetic part of the pertinent Dirac action for generic relativistic  fermions of (bare) mass $m$  
in the background (\ref{recoil}) 
reads: 
\bea\label{moddirac}
S &=& \int d^4 x \sqrt{g} ~ \overline{\psi}\Big[ \, ig_{\mu\nu}\gamma^\mu \partial^\nu  - m \Big] \, \psi = \nonumber \\
&=& \int d^4 x \Big(1 + u_i u_j \delta^{ij}  \Big) \overline{\psi} \Big[ i\eta_{\mu\nu}\gamma^\mu \partial^\nu  
+ iu_i\gamma^0 \partial^i  +  i\gamma^i u_i \partial_0 - m  \Big] \psi ~,
\eea
where we have taken into account that the determinant $g$ of the deformed metric is $(1 + u_i u_j \delta^{ij}) $,
and the recoil velocity is given by eq.(\ref{recoil}).

We may next consider an average over ensembles of D-particles in a situation where  the momentum transfer 
variable $r$ in (\ref{recoil}) is \emph{stochastic}~\cite{mavroLV} 
\be\label{stoch}
\ll r \gg = 0 ~, \quad \ll r^2 \gg  \ne 0 
\ee
In such stochastic ``foam'' situations, then, the effective mass scale $M$ (\ref{star}) 
can be understood as an average effect  
$M \sim M_s/g_s \sqrt{\ll (n^\star r)^2 \gg }$. 
Applying these ideas to the present case of fermion field propagation in such a stochastic D-particle medium, 
we may next sum over world-sheet genera, which represents quantum effects of the string. Such a summation implies the replacement 
of the momentum $p_i$ appearing in the definition (\ref{recoil}) of the recoil velocity by \emph{a quantum mechanical operator }
\begin{equation}\label{quant}
p_i \rightarrow -i \widehat p_i 
\end{equation}
in units of $\hbar =1 $ we are working on. This implies that $u_i $ in (\ref{moddirac}) is now a derivative operator
and thus normal ordering ambiguities arise~\cite{mavroLV}. Specifically, one first writes 
\be\label{oprel}
\overline \psi (\dots u_i u_j \delta^{ij}  ) \psi = (1 - \tilde c) u_i u_j \delta^{ij} \overline{\psi } (\dots ) \psi 
+ \overline{\psi }\tilde c u_i u_j \delta^{ij} \psi ~, 
\ee
where $\tilde c$ some real number; upon the canonical quantisation (\ref{quant}) the recoil velocity (\ref{recoil}) becomes an operator, hence
$$ u_i u_j \delta^{ij} \rightarrow  (g_s \, r n^\star)^2 \frac{\Delta }{M^2_s}~, \quad \Delta \equiv \partial_i \partial_j \delta^{ij}~.$$ 
The parts of the operator $\Delta$ to the left of the fermion $\overline \psi$ in (\ref{oprel}) yield total spatial derivatives in the 
effective Lagrangian which are irrelevant in the flat Minkowski background space-time.

According to the above ordering prescription, then, upon taking 
into account the averaging (\ref{stoch}),  the action (\ref{moddirac}) becomes 
\be\label{finalaction}
S= \int d^4 x ~\overline{\psi} \Big[ i \gamma^\mu \partial_\mu \Big(1 - \frac{\tilde a\, \Delta}{M^2} \Big) - 
m \Big(1 - \frac{\tilde b\, \Delta}{M^2} \Big) \Big] \psi
\dots ~, \quad M = \frac{M_s}{g_s \, \sqrt{\ll (n^\star r)^2 \gg}} 
\ee
where the indices are now contracted with the flat Minkowski space-time background metric and the constants $\tilde a, \tilde b > 0$ represent 
quantum ordering ambiguities. One may consider further four-fermion contact interaction terms in (\ref{finalaction}), 
which may be derived in the low-energy limit by the exchange of heavy string states or interactions with the D-particles themselves. Such 
terms are also suppressed by the heavy effective scale $M$.

The effects of the D-particle medium are then hidden in the LIV terms suppressed by $M^2$. Eventually one takes $\ll (n^\star \, r)^2 \gg \,  \to \, 0$, 
and as such the Lorentz Invariant limit is restored. However, as we shall see, this `regulator' LIV has important consequences for the generation 
of masses for the fermions, even if the latter have weak four fermion interaction couplings. In this sense, the D-particle medium catalyses 
mass generation which otherwise would require 
strong couplings.

\section*{Appendix B:\\ Derivation of the kinetic term for the auxiliary field}

For simplicity, we neglect here higher order derivative terms, since these will only provide corrections of order $1/M$ in the kinetic term for 
the auxiliary field, and we consider the Lorentz symmetric case.\\
In order to derive the kinetic term for the auxiliary field, one needs to take a non-homogeneous configuration, and we consider the plane wave
\be
\phi=\phi_1+\rho\Big(\exp(ik_\mu x^\mu)+\exp(-ik_\mu x^\mu)\Big)~,
\ee
where $\rho<<\phi_1$. The integration over fermions leads to the formal expression
\be
i\mbox{Tr}\ln\left(i\slashed\partial-g\phi\right)~,
\ee
which is expanded in $\rho$ and $k$, in order to identify the kinetic term, which is proportional to $k^2\rho^2$.
We then first need to expand to the second order in $\rho$ the expression
\bea
&&\ln\left(\slashed p-g\phi\right)\\
&=&\ln\left[(\slashed p-g\phi_1)\delta(p+q)-g\rho\left(\delta(p+q+k)+\delta(p+q-k)\right)\right]\nn
&=&\delta(p+q)\ln(\slashed p-g\phi_1)-g\rho\frac{\slashed p+g\phi_1}{p^2-g^2\phi_1^2}\left(\delta(p+q+k)+\delta(p+q-k)\right)\nn
&&+\frac{g^2\rho^2}{2}\frac{\slashed p+g\phi_1}{p^2-g^2\phi_1^2}
\left(\frac{\slashed k-\slashed q-g\phi_1}{(k-q)^2-g^2\phi_1^2}+\frac{-\slashed k-\slashed q-g\phi_1}{(k+q)^2-g^2\phi_1^2}\right)~\delta(p+q)\nn
&&+\frac{g^2\rho^2}{2}\frac{\slashed p+g\phi_1}{p^2-g^2\phi_1^2}\frac{-\slashed k-\slashed q-g\phi_1}{(k+q)^2-g^2\phi_1^2}~\delta(p+q+2k)\nn
&&+\frac{g^2\rho^2}{2}\frac{\slashed p+g\phi_1}{p^2-g^2\phi_1^2}\frac{\slashed k-\slashed q-g\phi_1}{(k-q)^2-g^2\phi_1^2}~\delta(p+q-2k)
+{\cal O}(\rho^3)~,\nonumber
\eea
such that
\bea
&&\mbox{Tr}\ln\left(\slashed p-g\phi\right)\\
&=&V\mbox{tr}\int\frac{d^4p}{(2\pi)^4}\ln(\slashed p-g\phi_1)\nn
&&+V\frac{g^2\rho^2}{2}\mbox{tr}\int\frac{d^4p}{(2\pi)^4}
\frac{\slashed p+g\phi_1}{p^2-g^2\phi_1^2}
\left(\frac{\slashed k+\slashed p-g\phi_1}{(k+p)^2-g^2\phi_1^2}+\frac{-\slashed k+\slashed p-g\phi_1}{(k-p)^2-g^2\phi_1^2}\right)
+{\cal O}(\rho^3)~,\nonumber
\eea
where $V$ is the space time volume. In the previous equation,
the first term corresponds to corrections to the potential $V(\phi_1)$, that we have already calculated and we therefore discard here. 
The second term is expanded in $k$ to give (ignoring higher orders in $\rho$)
\bea
&&\mbox{Tr} \ln\left(\slashed p-g\phi\right)\\
&=&2Vg^2\rho^2\int\frac{d^4p}{(2\pi)^4}\left(\frac{4(pk)^2}{(p^2-g^2\phi_1^2)^3}-\frac{2k^2}{(p^2-g^2\phi_1^2)^2}\right)
+~\mbox{$k$-independent terms}~+~{\cal O}(k^4)~,\nonumber
\eea
where the $k$-independent terms correspond to corrections to the potential, arising from $\rho\ne0$, and therefore can be omitted here.
Using the property
\be
\int d^4p ~f(p^2) p^\mu p^\nu=\frac{\eta^{\mu\nu}}{4}\int d^4p~ p^2f(p^2)~,
\ee
we obtain (considering only the relevant terms proportional to $k^2$)
\be
\mbox{Tr} \ln\left(\slashed p-g\phi\right)=2Vk^2g^2\rho^2\int\frac{d^4p}{(2\pi)^4}~\frac{-p^2+2g^2\phi_1^2}{(p^2-g^2\phi_1^2)^3}~.
\ee
We then perform a Wick rotation, regulate the integral by $M$ and replace $g\phi_1$ by the dynamical mass $m_{dyn}$ to finally obtain
\bea
i\mbox{Tr} \ln\left(\slashed p-g\phi\right)&=&V\frac{k^2g^2\rho^2}{8\pi^2}\int_0^{M^2/m_{dyn}^2}xdx\frac{2+x}{(1+x)^3}\\
&\simeq&V\frac{k^2g^2\rho^2}{4\pi^2}\ln\left(\frac{M}{m_{dyn}}\right)~.\nonumber
\eea
The kinetic term we are looking for is 
\be
\int d^4x~\frac{Z}{2}\partial_\mu\phi\partial^\mu\phi=VZk^2\rho^2~,
\ee
such that the identification with $i\mbox{Tr} \ln\left(\slashed p-g\phi\right)$ finally leads to
\be
Z=\frac{g^2}{4\pi^2}\ln\left(\frac{M}{m_{dyn}}\right)~.
\ee
We note here that this expression can also be obtained by the Bethe-Salpether approach, consisting in deriving the propagator for the 
bound state $\ol\psi\psi$ \cite{bardeen}. 
As explained in subsection 2.3, the Lorentz symmetric limit (\ref{gpropto}) freezes the auxiliary field to its vev, since, for a fixed dynamical mass,
the kinetic term vanishes in this limit
\be
\lim_{M\to\infty}Z\propto\lim_{M\to\infty}\frac{1}{M}\ln\left(\frac{M}{m_{dyn}}\right)=0~.
\ee

\end{document}